\documentclass[letter,twocolumn]{jpsj3}
\usepackage{txfonts}
\usepackage{color}
\usepackage{amsmath}

\newcommand{\GdRuAl}{Gd$_3$Ru$_4$Al$_{12}$}

\title{Helical Ordering of Spin Trimers in a Distorted Kagome Lattice of \GdRuAl\ Studied by Resonant X-ray Diffraction}

\author{Takeshi Matsumura$^{1}$, Yusaku Ozono$^{1}$, Shintaro Nakamura$^{2}$, Noriyuki Kabeya$^{3,4}$, and Akira Ochiai$^{4}$ }

\inst{$^{1}$Department of Quantum Matter, AdSM, Hiroshima University, Higashi-Hiroshima, Hiroshima 739-8530, Japan \\
$^{2}$Institute for Materials Research, Tohoku University, Sendai 980-8577, Japan \\
$^{3}$Institute for Excellence in Higher Education, Tohoku University, Sendai 980-8576, Japan\\
$^{4}$Department of Physics, Graduate School of Science, Tohoku University, Sendai 980-8578, Japan 
}

\abst{Successive magnetic phase transitions at $T_1$=17.5 K and $T_2$=18.5 K in \GdRuAl, with a distorted kagome lattice of Gd ions, 
is studied using resonant X-ray diffraction with polarization analysis. 
It has been suggested that in this compound the $S=7/2$ spins on the nearest-neighbor Gd-triangle form a ferromagnetic trimer and the Gd lattice can be effectively considered as an antiferromagnetic triangular lattice of $S=21/2$ spin trimers [S. Nakamura \textit{et al.}, Phys. Rev. B \textbf{98}, 054410 (2018)].  
We show that the magnetic order in this system is described by an incommensurate wave vector $\mib{q}\sim\;$(0.27, 0, 0), which varies slightly with temperature. 
In the low temperature phase below $T_1$, the experimental results are well explained by considering that the spin trimers form a helical order with both the $c$-axis and $c$-plane components. In the intermediate phase above $T_1$, the $c$-axis component vanishes, resulting in a sinusoical structure within the $c$-plane. 
The sinusoidal-helical transition at $T_1$ can be regarded as an ordering of chiral degree of freedom, which is degenerate in the intermediate phase.  
}

\begin{document}
\maketitle
Nontrivial orderings of spins in frustrated magnetic systems have attracted a long-standing interest. 
Various kinds of non-collinear or non-coplanar orderings, as well as successive orderings through partially disordered states, which take place to release the frustration, have been widely studied both experimentally and theoretically, especially in triangular, kagome, and pyrochlore lattice systems\cite{Kawamura98,Miyashita85,Kadowaki87,Inami09,Mitsuda00,Nakajima07,Nishiwaki17,Motoya18}. 
The chiral degree of freedom often plays an important role in providing an understanding of the nontrivial nature of the ordered states\cite{Kawamura98,Miyashita85}. 
In the present letter, we report a novel case realized in a distorted kagome lattice of $S$=7/2 $4f$-spins of Gd ions, which can effectively be considered as a triangular lattice of trimerized $S$=21/2 spins. 

R$_3$Ru$_4$Al$_{12}$ (R=rare earth), with a distorted kagome lattice of R ions  (hexagonal space group $P6_3/mmc$), is a metallic system that has been attracting significant interest in recent years because of its characteristic lattice structure and the possibilities of phenomena originating from geometrical frustration\cite{Troc12,Nakamura14,Nakamura15,Gorbunov14,Gorbunov16,Gorbunov18,Ishii18,Chandragiri16}.  
With respect to \GdRuAl, it was recently pointed out by Nakamura \textit{et al.} that the three Gd$^{3+}$ spins of $S$=7/2 on the nearest neighbor triangle form a spin trimer with $S$=21/2\cite{Nakamura18}. 
They showed that the temperature dependence of the magnetic susceptibility and specific heat, as well as the magnetic entropy, can be well explained by an isolated trimer model expressed by $\mathcal{H}=J(\mib{S}_1\cdot\mib{S}_2 + \mib{S}_2\cdot\mib{S}_3 + \mib{S}_3\cdot\mib{S}_1)$ with a ferromagnetic exchange constant of $J$=13.5 K. 
The first excited state of $S$=19/2 is located at 142 K in this model. 
The agreement between the experiment and the calculation supports the validity of the spin-trimer model. 
In addition, they found that \GdRuAl\ exhibits successive phase transitions at $T_1$=17.5 K and $T_2$=18.5 K. In the following paper, it is interpreted that the $c$-plane and $c$-axis components order at $T_2$ and $T_1$, respectively\cite{Nakamura19}. 
It is also speculated that the magnetic structure below $T_2$ is of a non-collinear type, which consists of both the $c$-axis and $c$-plane components. 

\begin{figure}
\begin{center}
\includegraphics[width=8cm]{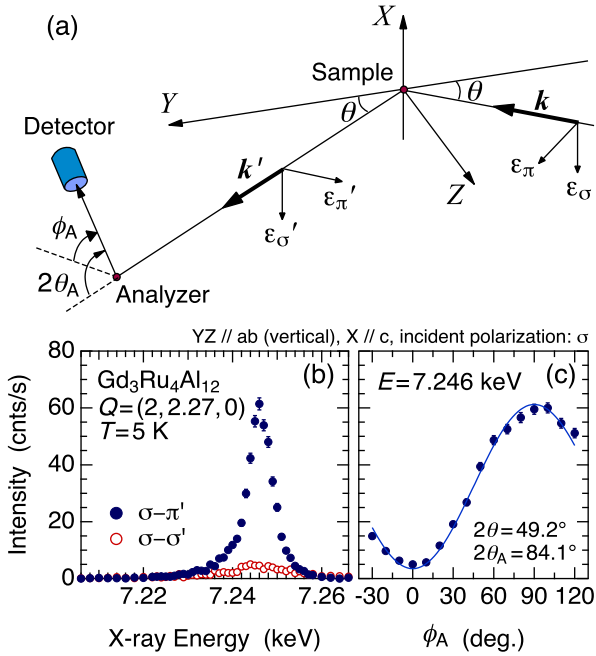}
\caption{(Color online) (a) Scattering configuration of the experiment. We take the $Z$-axis along the scattering vector $\mib{Q}=\mib{k}'-\mib{k}$, $X$-axis along $\mib{k}\times\mib{k}'$, and $Y$-axis along $\mib{k}+\mib{k}'$. 
The relation between the $XYZ$-coordinate and the crystal axes are shown in each figure. 
(b) X-ray energy dependences of the intensity of the (2, 2.27, 0) peak for $\sigma$-$\pi'$ ($\phi_{\text{A}}=90^{\circ}$) and $\sigma$-$\sigma'$ ($\phi_{\text{A}}=0^{\circ}$). 
(c) Analyzer angle ($\phi_{\text{A}}$) dependence of the resonant intensity at 7.246 keV. 
The solid line is a fit by assuming a scattering amplitude with $F_{\sigma\pi'} \neq 0$ and $F_{\sigma\sigma'}=0$. 
 }
\label{fig:configEscanPol}
\end{center}
\end{figure}

In order to investigate the magnetic structure of \GdRuAl, especially to focus on the trimer formation and the successive phase transition, we utilize resonant X-ray diffraction, which has an advantage over neutron diffraction since Gd is a strong neutron-absorbing element. 
Resonant X-ray diffraction experiments were performed at BL-3A of the Photon Factory, KEK, Japan. 
We first used a four-circle diffractometer with a vertical scattering plane (incident polarization: $\sigma$) to search for the magnetic signals, and subsequently used a two-axis diffractometer with a horizontal scattering plane (incident polarization: $\pi$), equipped with a vertical field 8 Tesla cryomagnet, to perform more detailed polarization analyses and investigate the magnetic field effects. The scattering configuration is shown in Fig.~\ref{fig:configEscanPol}(a). 
We used X-ray energies near the $L_3$ absorption edge of Gd and used a Cu-220 crystal analyzer to analyze the polarization of the diffracted beam. 
The method of sample preparation and the basic properties are described in Ref.~\citen{Nakamura18}. 
We used a plate-shaped single-crystalline sample (0.45 mm thick and $1.7\times 2.5$ mm$^2$ area) with a flat and mirror polished $a$-plane $(2\bar{1}\bar{1}0)$ surface. 

In our first experiment using the four-circle diffractometer, we found magnetic signals at incommensurate wave vectors corresponding to $\mib{q}$ = (0.27, 0, 0) at the lowest temperature of 5 K. 
Figure~\ref{fig:configEscanPol}(b) shows the X-ray energy dependences of the (2, 2.27, 0) peak for $\sigma$-$\pi'$ and $\sigma$-$\sigma'$ scattering configurations. The intensity for $\sigma$-$\pi'$ exhibits a resonant peak at 7.246 keV, corresponding to the $E1$ resonance between the $2p_{3/2}$ and $5d$ states of Gd. 
This signal contains information of the spin configuration of Gd-$4f$ electrons. 
The detailed $\phi_{\text{A}}$ dependence of the intensity is shown in Fig.~\ref{fig:configEscanPol}(c), which is well explained by a scattering amplitude with $F_{\sigma\pi'} \neq 0$ and $F_{\sigma\sigma'}=0$, as shown by the solid line. 
The non-vanishing intensity at $\phi_{\text{A}}=0^{\circ}$ ($\sigma$-$\sigma'$) is due to the contamination from the $\sigma$-$\pi'$ scattering resulting from the condition $2\theta_{\text{A}}\neq 90^{\circ}$, which is included in the fitting curves throughout the paper.  
Since the scattering amplitude for the $E1$ resonance originating from the magnetic dipole moment $\mib{m}$ is proportional to 
$(\mib{\varepsilon}' \times \mib{\varepsilon})\cdot \mib{m}$, $F_{\sigma\sigma'}$ vanishes, whereas  $F_{\sigma\pi'}$ remains finite\cite{Hannon88}. 
The experimental result shows that the resonant signal is of magnetic dipole origin.  
The higher harmonic peaks of $2\mib{q}$ and $3\mib{q}$ with intensities larger than 1/20 of the main peak were not detected in the present study.

\begin{figure}
\begin{center}
\includegraphics[width=8cm]{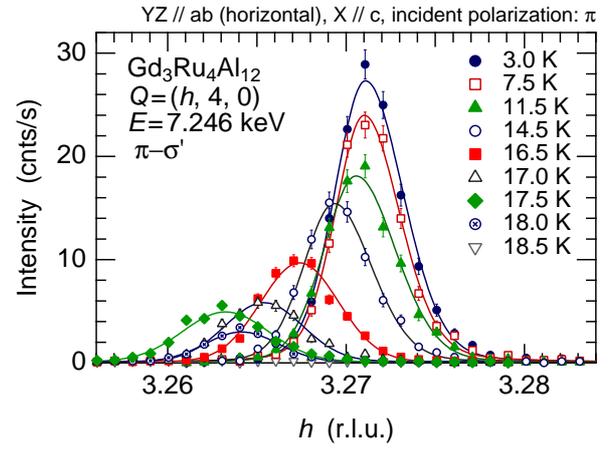}
\caption{(Color online) Temperature dependence of the (3, 4, 0)+$(q, 0, 0)$ peak profile measured at the resonance energy of 7.246 keV for $\pi$-$\sigma'$. The solid lines are the fitted profiles. 
 }
\label{fig:hscans0Tps}
\end{center}
\end{figure}

Figure \ref{fig:hscans0Tps} shows the temperature dependence of the scan profile along the (3, 4, 0)+$(q, 0, 0)$ line in the reciprocal space. 
It is noted that the measurement was performed using the two-axis diffractometer. 
The peak intensity decreases with increasing $T$, reflecting the reduction in the ordered magnetic moments. 
In addition, interestingly, the peak position shifts with $T$, and above $T_1$=17.5 K, it begins to shift to the opposite direction.

\begin{figure}
\begin{center}
\includegraphics[width=8.5cm]{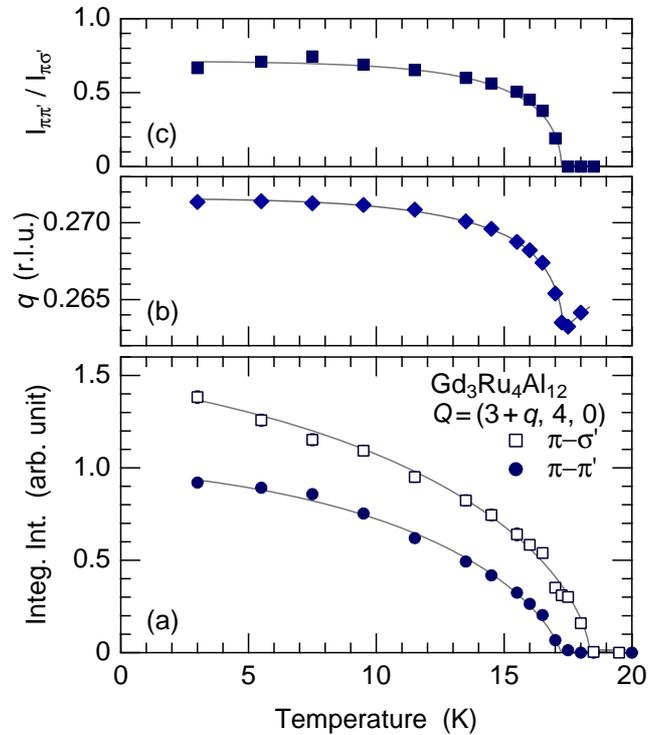}
\caption{(Color online) 
(a) Temperature dependence of the integrated intensity of the (3, 4, 0)+$(q, 0, 0)$ peak-profile for $\pi$-$\sigma'$ and $\pi$-$\pi'$ processes. 
(b) Temperature dependence of the $q$ value. 
(c) Temperature dependence of the ratio of the $\pi$-$\pi'$ intensity to the $\pi$-$\sigma'$ intensity. 
The solid lines are guides for the eye. 
}
\label{fig:TdepParams0T}
\end{center}
\end{figure}

The parameters obtained from the measurement are displayed in Fig.~\ref{fig:TdepParams0T}. 
As shown in Fig.~\ref{fig:TdepParams0T}(a), the $\pi$-$\pi'$ intensity vanishes at $T_1$=17.5 K, whereas the $\pi$-$\sigma'$ intensity vanishes at $T_2$=18.5 K. This difference clearly corresponds to the successive transition reported by Nakamura \textit{et al.} in Ref.~\citen{Nakamura18}. 
From the scattering amplitude $(\mib{\varepsilon}' \times \mib{\varepsilon})\cdot \mib{m}$, we see that 
the $\pi$-$\pi'$ intensity reflects the magnetic component perpendicular to the scattering plane, whereas the $\pi$-$\sigma'$ intensity reflects the parallel component. 
In the present geometry, the $\pi$-$\pi'$ and $\pi$-$\sigma'$ intensities reflect the $c$-axis and $c$-plane components, respectively. 
The difference in the transition temperature, therefore, directly shows that the $c$-plane component orders at $T_2$ and then the $c$-axis component orders at $T_1$. 

Figure~\ref{fig:TdepParams0T}(b) shows the temperature dependence of the $q$-value. 
It decreases with increasing $T$, which seems almost proportional to the variation of the magnetic order parameter. 
Above $T_1$, interestingly, the $q$-value begins to increase in the opposite direction. 
A similar behavior is also observed in GdNi$_2$B$_2$C\cite{Detlefs96}. 
We refer to Fig.~\ref{fig:TdepParams0T}(c) later.

\begin{figure}
\begin{center}
\includegraphics[width=8cm]{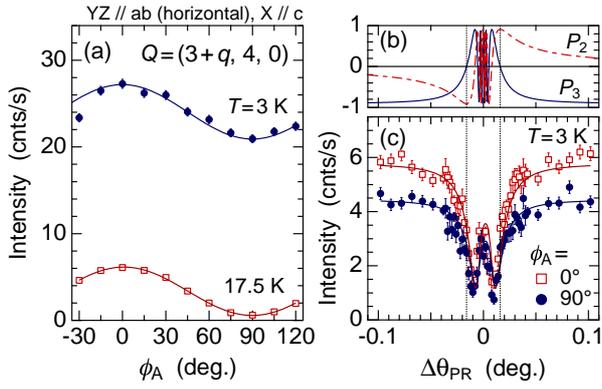}
\caption{(Color online) 
(a) Analyzer angle ($\phi_{\text{A}}$) dependence of the (3, 4, 0)+$(q, 0, 0)$ peak-intensity at 3 K and 17.5 K. 
(b) Offset angle $(\Delta \theta_{\text{PR}})$ dependence of the Stokes parameters of the incident X-ray after transmitting the phase retarder. 
$P_2$ and $P_3$ represent the degree of circular (right handed, $+1$, or left handed, $-1$) and linear ($\sigma$, $+1$, or $\pi$, $-1$) polarizations, respectively. 
The dotted vertical lines represent the positions of the circularly polarized states. The beam is depolarized around $\Delta \theta_{\text{PR}}=0^{\circ}$. 
(c) Offset angle dependence of the (3, 4, 0)+$(q, 0, 0)$ peak-intensity measured for the analyzer angles at $0^{\circ}$ and $90^{\circ}$. 
The solid lines in (a) and (c) are calculations assuming a helical trimerized magnetic structure described in the text. 
}  
\label{fig:PolAnaPR}
\end{center}
\end{figure}

In the geometry of the horizontal scattering plane, both the $\pi$-$\sigma'$ and $\pi$-$\pi'$ scattering amplitudes are involved in the scattering process. 
This coexistence could provide important knowledge that cannot be obtained in the geometry with $\sigma$ incident polarization. Figure~\ref{fig:PolAnaPR}(a) shows the $\phi_{\text{A}}$ dependence of the peak intensity measured at 3 K, the lowest temperature, and at 17.5 K, where the $c$-axis component has vanished. 
The data at 3 K exhibits a moderate oscillation without vanishing, indicating that the moments are non-collinear. 
To obtain more information on the magnetic structure, we inserted a diamond phase retarder system in the incident beam and measured the variation in intensity by changing the degree of linear and circular polarizations of the incident beam. By rotating the angle $\theta_{\text{PR}}$ of the diamond crystal around the 111 Bragg-angle $\theta_{\text{B}}$, where the scattering plane is tilted by 45$^{\circ}$, a phase difference arises between the $\sigma$ and $\pi$ components. 
By manipulating the offset angle $\Delta \theta_{\text{PR}}=\theta_{\text{PR}} - \theta_{\text{B}}$, we can tune the mixing ratio between the linear and circular polarization states, which is shown in Fig.~\ref{fig:PolAnaPR}(b) by using the Stokes parameters $P_2$ and $P_3$\cite{Matsumura17a}.

If the magnetic structure possesses some helicity, as in helical or cycloidal structures, asymmetrical $\Delta \theta_{\text{PR}}$ dependence is expected to be observed because of the different scattering cross-sections with respect to the helicity of the incident beam\cite{Matsumura17a,Matsumura17b}. 
However, the result shown in Fig.~\ref{fig:PolAnaPR}(c) is almost symmetric and it seems difficult to associate this result with helical or cycloidal structures. 
One possibility could be that structures with opposite helicity coexist and form equally populated domains. 
Although this is allowed in principle because the crystal structure of \GdRuAl\ is not chiral, the assumption of equal population seems too simplistic considering the appearance of unequal cycloidal domain populations in GdRu$_2$Al$_{10}$, whose crystal structure is not chiral as well\cite{Matsumura17a}. 
This point will be discussed later. 

\begin{figure}
\begin{center}
\includegraphics[width=8cm]{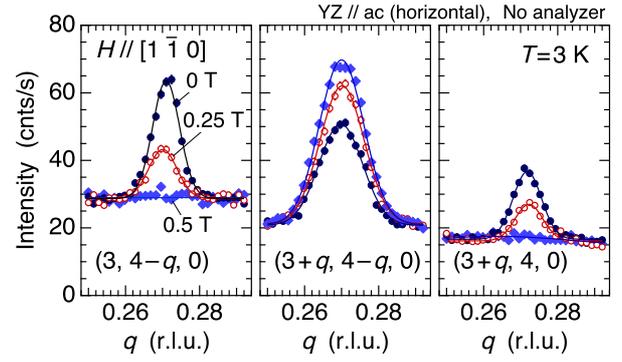}
\caption{(Color online) 
Magnetic field dependence of the peak profile corresponding to the three equivalent propagation vectors 
$(0,-q,0)$, $(q,-q,0)$, and $(q,0,0)$ around (3, 4, 0). The field is applied along $[1 \bar{1} 0]$. 
}
\label{fig:HabInt}
\end{center}
\end{figure}

To investigate the possibility of a multi-$q$ structure composed of equivalent propagation vectors among $(q, 0, 0)$, $(0, q, 0)$, and $(q, -q, 0)$, we applied a magnetic field along the $[1 \bar{1} 0]$ axis and measured the change in peak intensity corresponding to the three vectors. 
As clearly shown in Fig.~\ref{fig:HabInt}, the $(0,-q,0)$ and $(q,0,0)$ peaks soon disappear at 0.5 T, whereas the $(q,-q,0)$ peak survives and increases its intensity. 
This shows that the $(q,-q,0)$ single domain state is realized at 0.5 T and that the magnetic field prefers the magnetic domain whose propagation vector is parallel to the field direction. 
This means that the Fourier component $\mib{m}_{\mib{q}}$ is perpendicular to $\mib{q}$. 
We can also conclude that the magnetic structure at zero field is described by a single-$q$ component.

\begin{figure}
\begin{center}
\vspace{10mm}
\includegraphics[width=8.5cm]{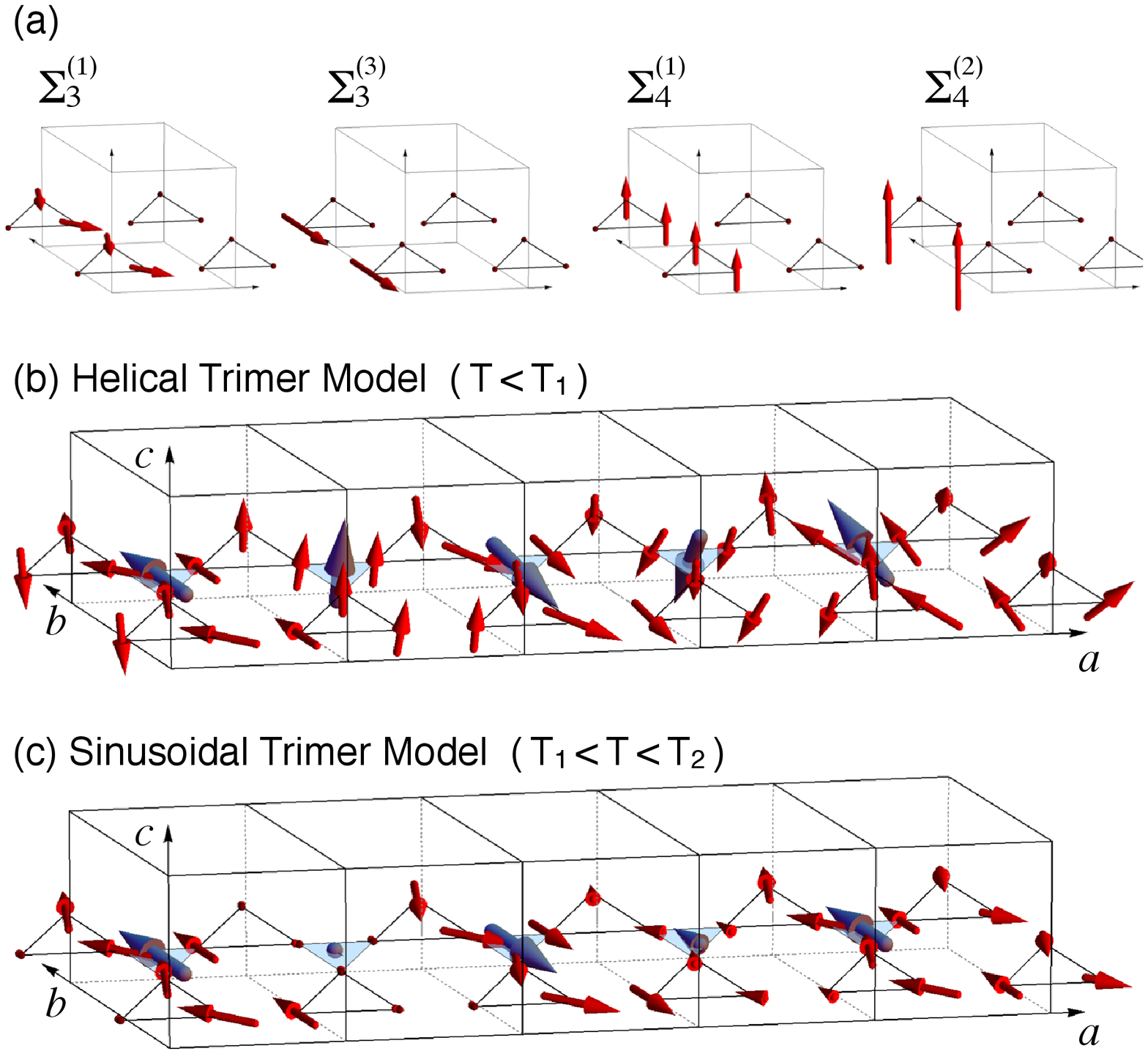}
\caption{(Color online) 
(a) Four types of irreducible representations of Gd spin structure with a propagation vector $(q, 0, 0)$ (see suppl. mat. for other representations). 
(b) Helical magnetic structure model with trimerized Gd spins on the colored triangles, propagating along the $a^*$-axis with $\mib{q}=(0.27, 0, 0)$ rotating in a clockwise direction. The total spin of each trimer, represented by the black arrow at the center of each triangle, is perpendicular to $\mib{q}$. 
(c) Sinusoidal trimer model without the $\Sigma_4$ component. Some trimers have very small and even vanishing ordered moments. }
\label{fig:magstModel}
\end{center}
\end{figure}

To explain the above experimental results, we introduce a helical trimer model, which is based on the irreducible representation of the incommensurate propagation vector $(q, 0, 0)$. 
Since there are six Gd ions in a unit cell, there arises 18 independent basis structures, which are classified into four irreducible representations\cite{SupplMat1}. 
$\Sigma_1$ and $\Sigma_3$ are related to the $c$-plane component, and $\Sigma_2$ and $\Sigma_4$ are related to  the $c$-axis component. 
Half of them are for Gd ions on the $z=1/4$ layer and the other half are for those on the $z=3/4$ layer. Since the two layers are independent in the representation, we consider the spin structure on the $z=1/4$ layer only. 
The same discussion can be applied to the $z=3/4$ layer as well. Four types of representations among the 18 are illustrated in Fig.~\ref{fig:magstModel}(a). 

A guiding principle to construct a model structure is to form a ferromagnetically coupled structure among the three spins on the nearest neighbor sites and make the total spin as large as possible. 
This is a reasonable assumption that is consistent with the interpretation of magnetic susceptibility and specific heat as the formation of spin trimer\cite{Nakamura18}. 
We also need to include both the $c$-plane and $c$-axis components. 
These requirements can be achieved by combining the four elements $\Sigma_1^{(2)}$, $\Sigma_1^{(3)}$, $\Sigma_4^{(1)}$, and $\Sigma_4^{(2)}$. 
Another possibility can be a combination of $\Sigma_3^{(1)}$, $\Sigma_3^{(3)}$, $\Sigma_4^{(1)}$, and $\Sigma_4^{(2)}$, which are shown in Fig.~\ref{fig:magstModel}(b). 
In the former model, the total moment of the trimer rotates in a cycloidal way, whereas in the latter model, it rotates in a helical way. 

By selecting the phase factors appropriately, a helical magnetic structure illustrated in Fig.~\ref{fig:magstModel}(b) is constructed, where the nearest-neighbor Gd spins are oriented in the same direction and rotate in a clockwise direction. 
By setting the amplitude of $\Sigma_4$ ($c$-axis component) 0.73 times as that of $\Sigma_3$ ($c$-plane component), the calculated curves in Figs.~\ref{fig:PolAnaPR}(a) and (c) are obtained, which well reproduce the experimental data. 
The smaller amplitude for $\Sigma_4$ means that the helical structure shown in Fig.~\ref{fig:magstModel}(b) is ellipsoidal.  
We assumed equal populations for the clockwise (cw) and counter-clockwise (ccw) helical domains in the calculation. 
The difference in the calculated intensity between cw and ccw is not so significant in the helical model for $(3+q, 4, 0)$, which means that the domain imbalance, if any, does not give rise to a significantly asymmetric intensity in the $\Delta\theta_{\text{PR}}$ scan of Fig.~\ref{fig:PolAnaPR}(c)\cite{SupplMat2}.  
We need to check other reflections to further confirm the model. 
It is  noted that these data cannot be reproduced by the cycloidal model using $\Sigma_1$ and $\Sigma_4$\cite{SupplMat2}.

The magnetic structure in the intermediate phase ($T_1 < T < T_2$) is illustrated in Fig.~\ref{fig:magstModel}(c), which is obtained from the $\Sigma_3$ component only. The $c$-axis component is zero as concluded from the vanishing $\pi$-$\pi'$ intensity in Figs.~\ref{fig:TdepParams0T}(a) and \ref{fig:PolAnaPR}(a). 
This is a sinusoidal structure of the trimers, whose directions are perpendicular to the propagation vector. 
The result that the magnetic moments first order at $T_2$ within the $c$-plane with decreasing $T$, and that the structure in the low-$T$ phase is slightly ellipsoidal with shorter moments along the $c$-axis, is consistent with the weakly anisotropic magnetic susceptibility with preferable orientation within the $c$-plane. 

One of the important aspects of the sinusoidal structure is that there remain magnetic sites with small, or even vanishing, ordered moments. 
This means that there remain unreleased magnetic entropy, or degeneracy, which need to be lifted at lower temperatures. 
In the present case of \GdRuAl, the degeneracy is lifted by the formation of a helical magnetic structure below $T_1$ by inducing the $c$-axis component, i.e., by transforming the structure into a non-collinear form.  
We remark that the helical structure of Fig.~\ref{fig:magstModel}(b) is chiral, whereas the sinusoidal structure of Fig.~\ref{fig:magstModel}(c) is collinear and not chiral. 
Therefore, the transition at $T_1$ can also be regarded as an ordering of the chiral degree of freedom, which is degenerate above $T_1$. 
In order to induce full magnetic moment at the small moment site in the sinusoidal structure, the moment needs to be induced along the $c$-axis, which breaks the chiral symmetry preserved in the intermediate phase. 
The temperature dependence of $I_{\pi\pi'}/I_{\pi\sigma'}$ in Fig.~\ref{fig:TdepParams0T}(c) is, in this sense, equivalent to that of the chiral order paramter. 
This situation has been discussed recently by Nakamura \textit{et al.} in Ref.~\citen{Nakamura19}, where the authors described the non-collinear structure as a T-shaped structure, i.e., it consists of both the $c$-axis and $c$-plane components.
Although the magnetic structure discussed in Ref.~\citen{Nakamura19} is different from that of Fig.~\ref{fig:magstModel}(b), 
the idea captures the essence of the structure, i.e., it is non-collinear and non-coplanar, and is different from the 120$^{\circ}$ structure, which is the standard structure in a triangular lattice system such as CsNiCl$_3$ and GdPd$_2$Al$_3$\cite{Kadowaki87,Inami09}. 
A similar helical structure propagating along the $c$-plane is found in the high field phase of CuFeO$_2$ with ferroelectric polarization\cite{Mitsuda00,Nakajima07}.

In Gd compounds with weak crystal field anisotropy because of vanishing orbital moment, it is likely that the ordered structure just below the transition temperature reflects the most preferable state for the intrinsic magnetic interaction, which is not affected by the crystal field anisotropy nor by the developed magnetic moments. 
In the present case, the magnetic dipole interaction is considered to prefer in-plane ordering, which has lower energy than the ordered state with the $c$-axis component\cite{Nakamura19}. 
The propagation vector $\mib{q}\sim (0.27, 0, 0)$, on the other hand, is probably associated with $\chi(\mib{q})$ for the conduction electron in this compound, which mediates the magnetic exchange interaction via the Ruderman-Kittel-Kasuya-Yosida (RKKY) mechanism.

Another important observation in this work is the temperature dependence of the $q$ value shown in Fig.~\ref{fig:TdepParams0T}(b). 
In the intermediate phase, the $q$ value decreases with decreasing $T$, whereas it reverses its changing direction and increases with decreasing $T$ below $T_1$. 
It is interesting that the same behavior is also observed in GdNi$_2$B$_2$C, where the parallel and perpendicular components with respect to the basal $c$-plane orders successively~\cite{Detlefs96}. 
The temperature dependence of $q$ below $T_1$ is similar to that of the magnetic order parameter, 
which implies that the development of the order parameter causes a modification in the RKKY interaction and changes the resultant $q$ value\cite{Matsumura17a,JM91,Feng13}. 

In summary, we have performed resonant X-ray diffraction experiment on \GdRuAl\ with a distorted kagome lattice of Gd, where successive antiferromagnetic transitions of spin trimer, composed of ferromagnetically coupled $S$=7/2 spins, have been suggested. 
We constructed a model structure using the irreducible representation of the observed propagation vector $\mib{q}=(0.27, 0, 0)$, so that the trimer formation and the results of polarization analysis are explained consistently. 
The possible structure in the low-$T$ phase is an ellipsoidal helical trimer and that in the intermediate phase is an in-plane sinusoidal trimer.
The sinusoidal-helical transition to release the magnetic entropy in the present case can be regarded as a chirality ordering. 

\paragraph{Acknowledgements} 
This work was supported by JSPS KAKENHI Grant number 18K187370A. 
The synchrotron experiments were performed under the approval of the Photon Factory Program Advisory Committee (No. 2018G039). 
This work was also supported by Chirality Research Center (Crescent) in Hiroshima University and JSPS Core-to-Core Program, A. Advanced Research Networks.

\renewcommand{\thefigure}{S\arabic{figure}}
\setcounter{figure}{0}

\begin{fullfigure}[t]
Supplemental Material\\

\vspace{10mm}
\includegraphics[width=8.5cm]{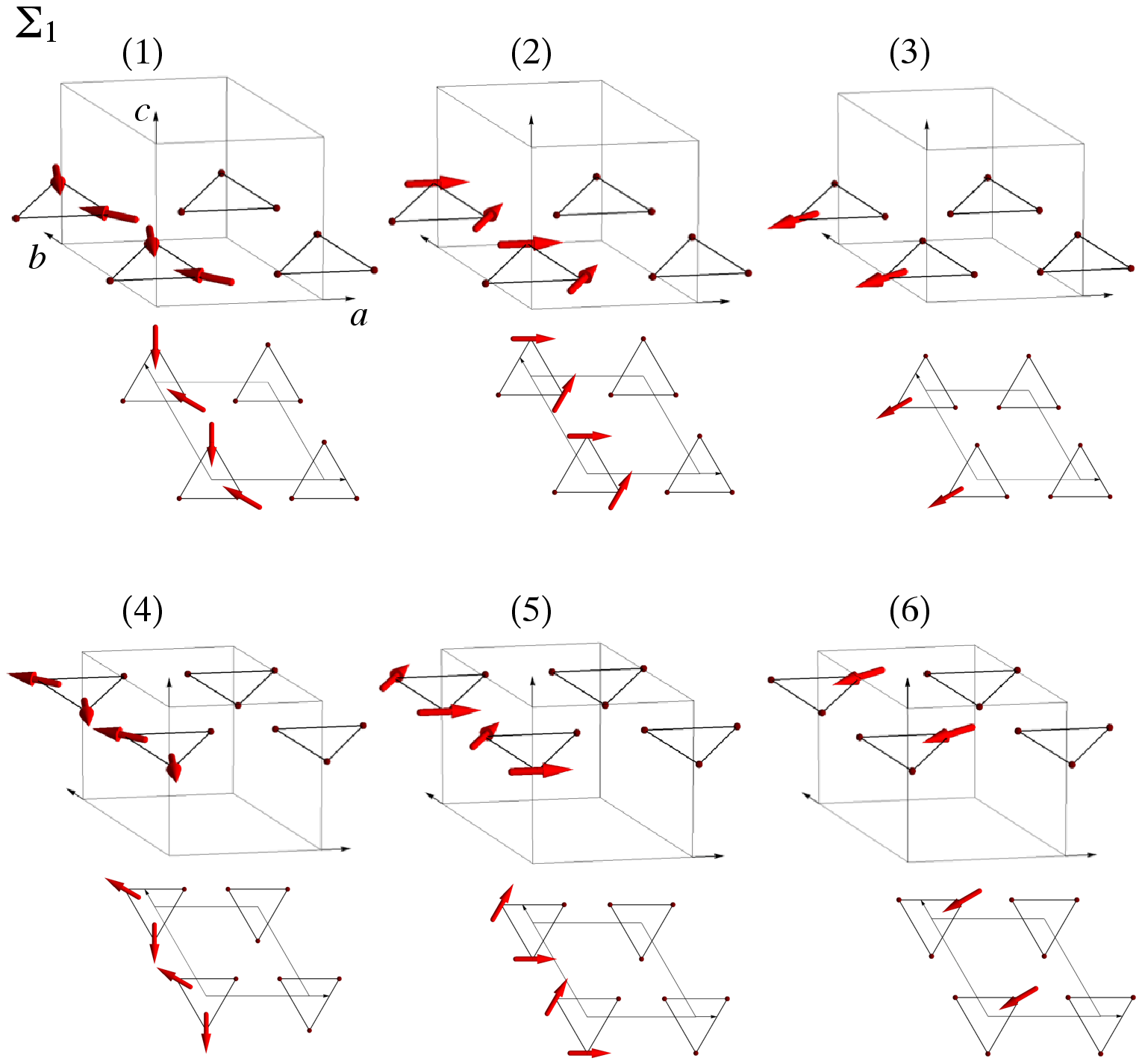}\hfill
\includegraphics[width=8.5cm]{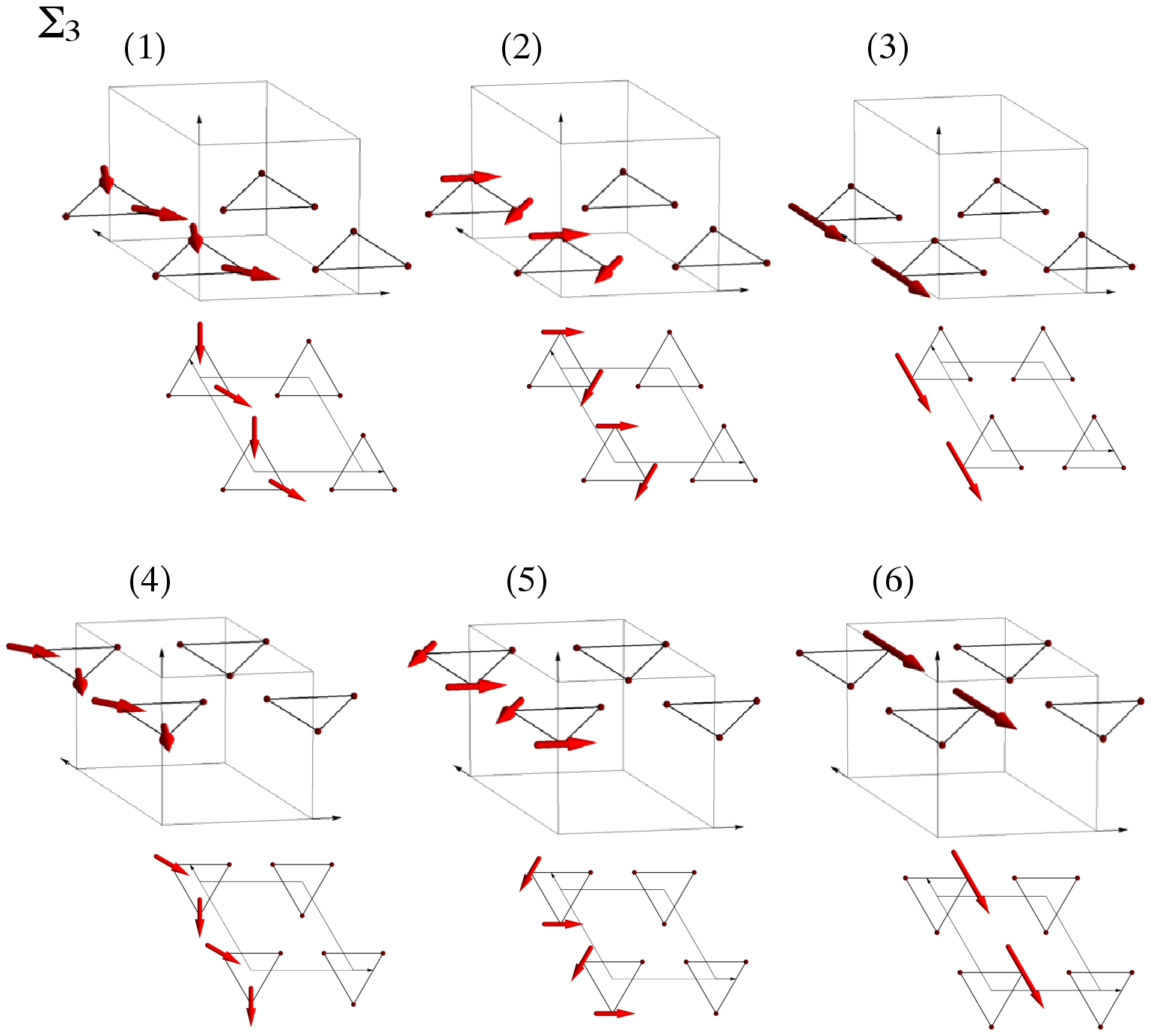}\\
\includegraphics[width=8.5cm]{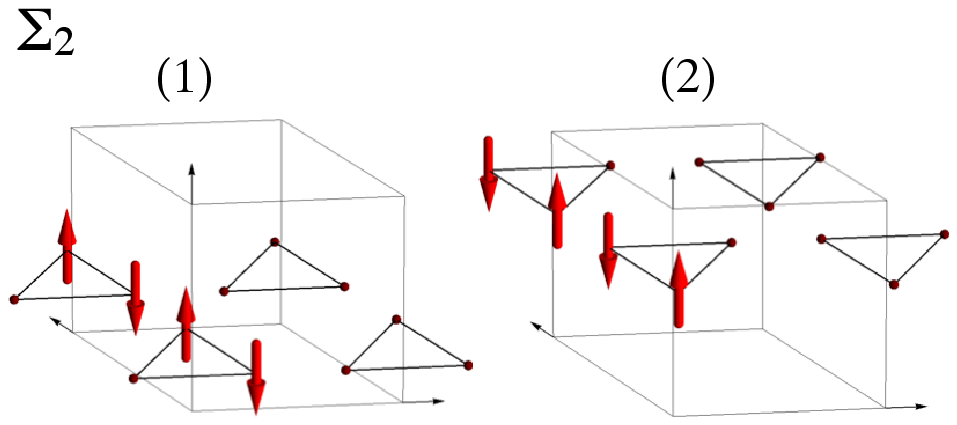}\hfill
\includegraphics[width=8.5cm]{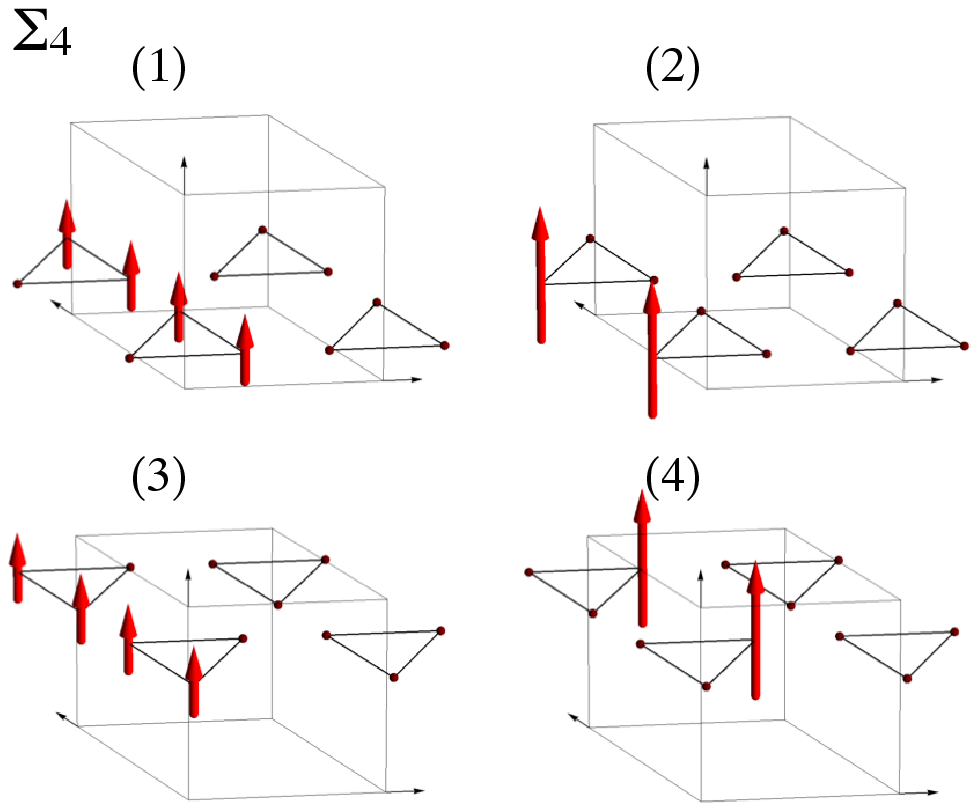}
\caption{Irreducible representations of the magnetic structure for the propagation vector ($k$, 0, 0) obtained by using the program TSPACE.}
\end{fullfigure}

\begin{fullfigure}[t]
Supplemental Material\\

\vspace{10mm}
\includegraphics[width=16cm]{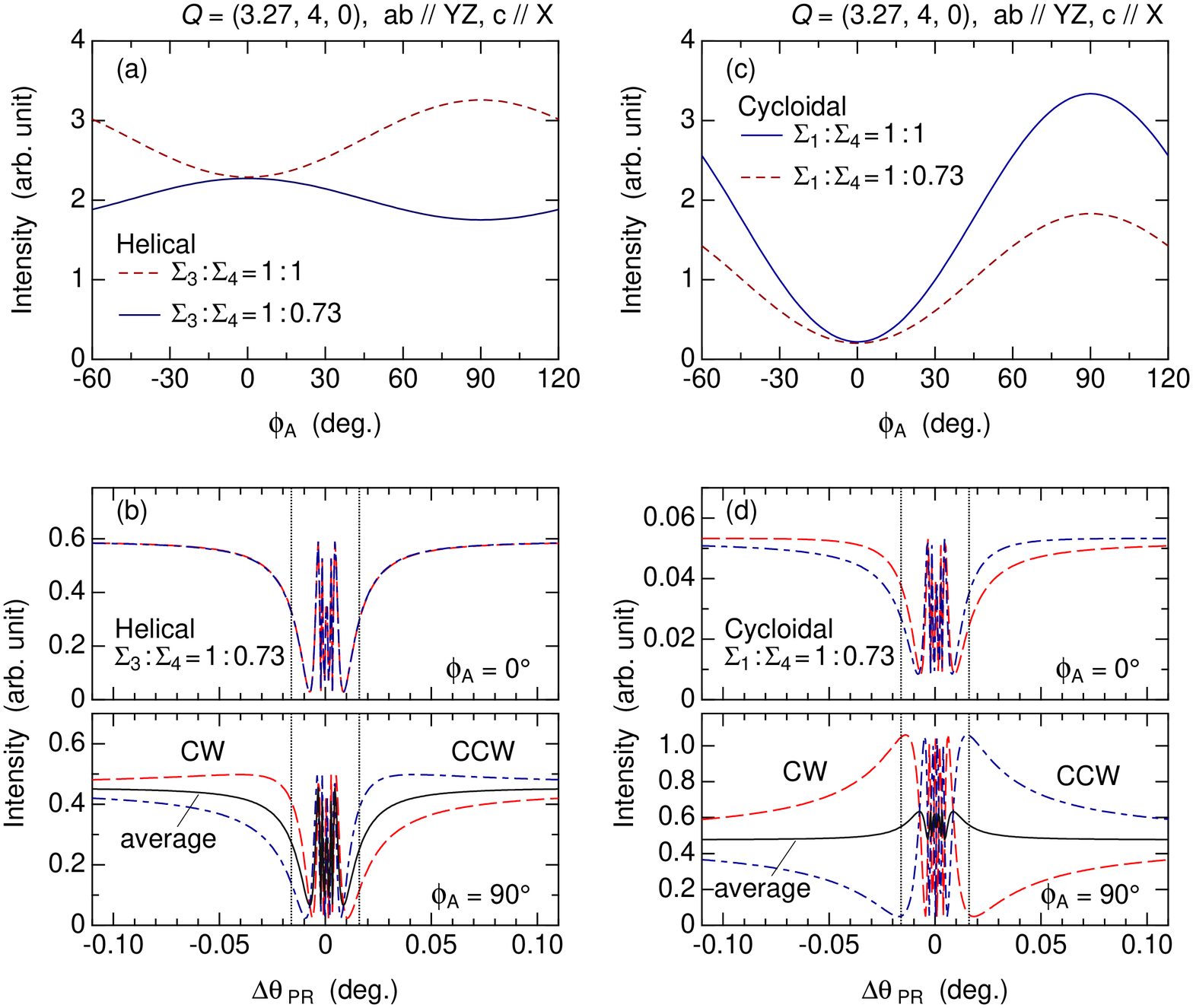}
\caption{Calculation of $\phi_{\text{A}}$ and $\Delta \theta_{\text{PR}}$ dependences shown in Fig. 4 for other model structures: cycloidal model and equal amplitude model.}
\end{fullfigure}

\end{document}